\documentclass[]{spie} 
\usepackage[]{graphicx}

\title{Test of Thick Pixelated Orbotech Detectors with and without 
Steering Grids}

\author{I. Jung\supit{a},  A. Garson III\supit{a}, H. Krawczynski\supit{a}, 
A. Burger\supit{b}, M. Groza\supit{b}, J. Matteson\supit{c}, R. T. Skelton\supit{c}
\skiplinehalf
\supit{a}Washington University in St. Louis, Department of Physics, 1 Brookings Dr., St. Louis, MO 63130\\
\supit{b}Fisk University, Department of Physics, 1000 Seventeenth Ave. North, Nashville, TN 37208\\
\supit{c}Center for Astrophysics and Space Sciences (CASS) - 0424
University of California, San Diego
9500 Gilman Drive
La Jolla, CA  92093-0424}
\authorinfo{Further author information: (Send correspondence to I.J.)\\I.J.: E-mail: jung@physics.wustl.edu, Telephone: 1 314 935 6254}

\begin{document} 
\maketitle 
  
\begin{abstract}
We report here on the optimization of 0.5~cm thick pixelated Orbotech CZT 
detectors with regards to the best contacting materials and the use of 
steering grids. We evaluated the performance of different contacting 
materials. Our study differs from earlier ones in that we investigated 
the performance of different anode and cathode materials separately. 
We obtain the best performance with Au cathodes. 
For different anode materials Ti and In give the best energy resolutions.
The detector (2.0$\times$2.0$\times$0.5~cm$^3$, 8$\times$8 pixels)
shows excellent 59 keV, 122 keV and 662 keV energy resolutions of 
1.4 keV, 1.9 keV, and 7.4 keV, respectively. 
Furthermore, we report on using steering grids to improve on the performance 
of the pixelated detectors. Previously, the benefit of steering grids had been 
limited by additional electronic noise associated with currents between the 
negatively biased steering grids and the anode pixels. 
We are currently exploring the possibility to isolate the steering grid 
from the CZT substrates by a thin layer of Al$_2$O$_3$. We performed a 
series of measurements to determine by how much the isolation layer reduces 
the grid-pixel currents. Comparing the currents between two Au contacts before
and after isolating one of the two contacts from the CZT with a 700~nm thick
layer of Al$_2$O$_3$, we measure that the isolation layer reduces 
the currents by a factor of about 10 at 500~V. We present some results from a detector before 
and after deposition of an isolated steering grid. The grid indeed improves on 
the detectors energy resolution and detection efficiency. 
We show that simulations can be used to model the anode to cathode 
charge correlation in excellent agreement with the experimental results.
\end{abstract}


\section{INTRODUCTION}
\label{sect:intro} 
Cadmium Zinc Telluride (CZT) has emerged as the detector material of choice 
for the detection of hard X-rays and soft gamma-rays with excellent position 
and energy resolution and without the need for cryogenic cooling. 
We report here on tests with CZT substrates from the company 
Orbotech Inc. \cite{Orbotech} (formerly Imarad).
They are grown 
with a Modified Horizontal Bridgeman (MHB) process. The MHB process results in a 
high yield of good crystals and in very uniform substrates.
\begin{figure}[bth] 
\vspace{-0.4cm}
\centering
\includegraphics[width=8cm]{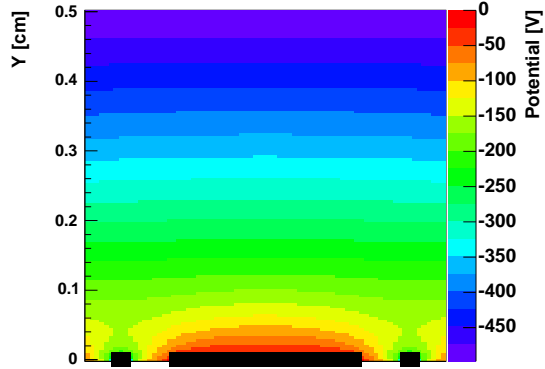}
\vspace*{-0.6cm}
\caption{Potential distribution of a central pixel in a 0.5~cm thick detector 
with pixel pitch of 0.24 mm, pixel width of 0.16 mm and a steering grid width 
of 0.016 mm from a 3-D detector simulation. While the anode pixels were held at 
ground, the cathode was biased at -500~V and the steering grid at -300~V.}
\label{potential}
\end{figure}
The bulk resistivity ($\sim$ $10^{10}$ ohm cm) is lower than that of
the more conventionally grown High-Pressure Bridgman CZT.
Using high-workfunction cathode materials that form blocking Shottky contacts 
on the $n$-type CZT substrates, the bias currents can be reduced by more 
than one order of magnitude so that the associated shot noise no longer 
deteriorate the energy resolution \cite{Vada,Henric04,Jung05}.
Different cathode materials have been studied extensively. In this contribution, 
we give a brief report on results obtained with different anode materials. 
A full description of this study will be given elsewhere (Jung et al., in preparation).

\begin{figure}[t]
\centering
\includegraphics[width=12.5cm]{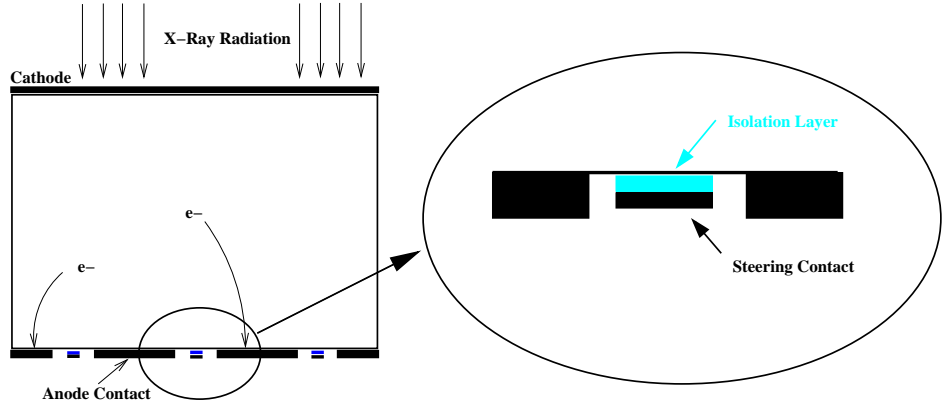}
\caption{Sketch of a CZT substrate with a steering grid isolated from the CZT
substrate by a high-resistivity film (dimensions not to scale). }
\label{fig:Layer}
\end{figure}

Most current CZT detectors use ``electron-only'' detection strategies \cite{Barret95,Luke95}
to ameliorate the problems associated with the poor hole mobilities and short
hole trapping times in CZT ($(\mu_h\:\tau_h \sim \rm(0.2-5)\cdot10^{-5}$cm$^2/V)$).
One can use for example the ``small pixel effect'' \cite{Barret95,Luke95} so that
charge generated inside the detector induces pixel currents only briefly before
impinging on the anode pixels. The anode signals become largely independent of 
the location of the charge generation and the ``depth of interaction'' (DOI).
An additional DOI correction can be used to further improve the energy
resolution. The DOI can be estimated from measuring the drift time of 
the electrons \cite{Kalem02,Zhang04} and from measuring the anode to cathode
charge ratio \cite{Kraw04}. Some groups (see e.g.\ \cite{Zhang04}) use steering 
grids to improve the performance of pixelated detectors. Steering grids ameliorate the 
energy resolutions and detection efficiencies of pixelated detectors by 
steering electrons away from the areas between pixels towards the pixels \cite{bolotnikov}.
Furthermore, steering grids modify the weighting potential of the pixels.
The result is a stronger small pixel effect and
an additional enhancement of the energy resolution of the detectors.
Most groups bias the steering grids at voltages of between -30 and -60~V 
relative to the anode pixels. Owing to the limited surface resistivity of CZT,
large grid-pixel currents start to flow for higher grid bias voltages.
However, simulations show that a grid bias between -100 V and -300 V is 
required to appreciably modify the electric field inside the detectors 
(see Fig.\ \ref{potential}). We are thus exploring the possibility to isolate 
the grids from the CZT substrates (Fig. \ref{fig:Layer}). The isolation 
layer can reduce the grid-pixel currents while allowing the electric field 
to enter the detector.
The second main topic of this contribution are studies of the effectiveness of 
Al$_2$O$_3$ isolation layers. Furthermore, we present some results from a 
detector before and after deposition of an isolated steering grid, and 
compare experimentally measured data with simulations.

\section{Detector Fabrication}
\label{sect:fabrication}
We use Orbotech CZT substrates, 2.0$\times$2.0$\times$0.5~cm$^3$, contacted with a 
planar cathode and 8$\times$8  anode pixels with a pitch of 0.25~cm 
and a pixel width of 0.16~cm. We polish the CZT substrates with Al suspension, and etch 
them in 5\% Br-95\% Methanol solution. After etching for 1.5~minutes, the substrates are 
rinsed in various Br-Methanol solutions with decreasing Br concentration. 
The contacts are deposited with an electron beam evaporator. 
We use masks and standard photolithographic techniques to 
deposit the pixels and/or the steering grids.
\subsection{Test Electronics}
The CZT detectors are temporarily mounted with gold-plated, spring-loaded ``pogo-pin'' contacts.
To measure the cathode-pixel 
and pixel-pixel IV curves 
a set-up consisting of a PC, a programmable Bertan high voltage power supply, a programmable AD switch, and a Keithley electrometer is used. 

The energy resolution is determined mostly 
for flood-illuminating the detectors
with X-rays. A hybrid electronic readout is used. The pulse shape is read out for four 
channels: the central anode pixels and the cathode.  These pixels are AC coupled and amplified 
by a fast Amptek 250 amplifier followed by a second  amplifier stage. 
After digitizing the amplified signals using a 500 MHz oscilloscope  they 
are sent  via an Ethernet connection to a PC. With this setup the electron 
drift-time can be determined with an accuracy of 10 ns. 
 Sixteen additional channels 
can be read out with an ASIC to measure the pulse height information. 
The ASIC gives amplified and shaped signals which we digitize with a custom 
designed VME board. The FWHM noise of both readout chains lies 
between 5 keV and 10 keV. 

We have made a few measurements using collimated X-rays. 
A 7.5 cm thick tungsten collimator is used with a tapered hole. 
The hole diameter is 0.02 cm at the source and 0.05 cm at the detector end of 
the collimator. The CZT cathode is situated 0.10 cm off the collimator.
The collimator position is controlled by a x-y stage with 1$\mu$m 
accuracy. Eight channels can be read out, the information obtained is the 
pulse height of the signals. The FWHM noise of anode channels 
lies between 5.75~keV to 8.1~keV for cathode biases between $-100$~V and 
$-1000$~V and grid bias between $-30$~V and $-120$~V. The FWHM noise of the 
cathode channel lies between 7.2~keV and 41~keV.
More information on this experimental setup as well as previous results
can be found in \cite{Kalem02}.

We measure the detector performance at different temperatures by placing the
detector and the readout electronics inside a Tenney climate chamber and cooling
or heating the set-up to the desired temperature.
\section{CZT detectors contacted  with various metals}
\label{sec:altcontacts}
\begin{figure}[b]
\centering
\includegraphics[width=2.5in]{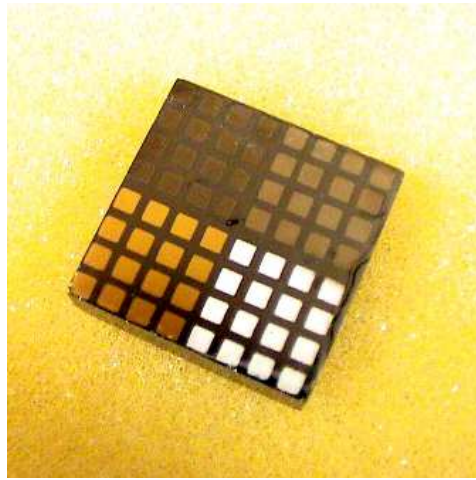}
\caption{Picture of a pixellated CZT detector (2 x 2 x 0.5 cm$^3$). 
The anode area was divided into four quadrants, and each quadrant was contacted 
with a different anode material (from top left clockwise, Ti, In, Au, Cr).}
\label{fig:contactmaterial}
\end{figure}
We have performed a detailed study of various contact metals by  optimizing 
the cathode and anode contacts separately.  In our earlier study \cite{Jung05} we used
Orbotech detectors contacted with In cathode and anodes. Replacing the  In cathodes with
Au, Cr, Ni, Pt and Ti cathodes, we observed a reduction of the cathode-pixel 
currents by factors of ten and higher. Whereas detectors with Au, Cr, In and Ti 
cathodes gave very similar 662 keV energy resolutions, Ni and Pt gave substantially 
poorer performance. In another study, we used identical materials for the 
cathode and anode contacts. Of Au, Cr, In and Pt, it was Au that gave the best result, 
followed in order of performance by Pt, In and Cr. 
In this study, we had used individual CZT crystals several times. 
This procedure eliminates uncertainties associated with a variable crystal 
quality that plague studies that use several crystals.
However, we observed that the performance of some crystals deteriorated with 
time, owing to the wear and tear from polishing off the contacts and
depositing new contacts. 

Here, we use a slightly different strategy, depositing four metals on four 
quadrants of one detector. The procedure has the advantage that the processing 
procedure is identical for all four metals. It has the disadvantage that the
crystal quality might vary across the crystal. We plan to exclude the latter
possibility by evaporating a single metal on the anode and test for
spatial variations of the crystal quality. Based on the previous results we used \begin{figure}
\vspace{-0.4cm}
\centering
\includegraphics[width=8cm]{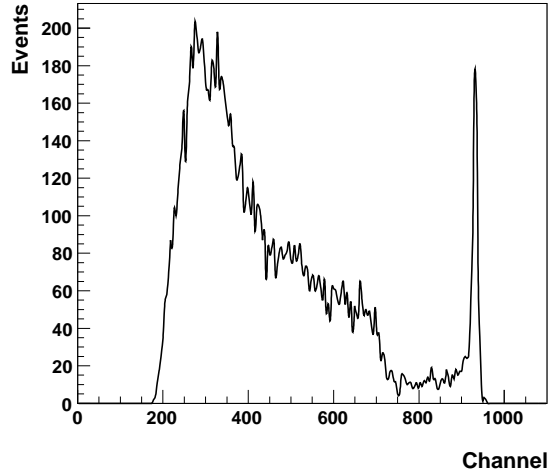}
\caption{662 keV energy spectra of an Orbotech detector contacted with Au cathode and four different anode materials. Shown is the spectra for a Ti pixel. The resolution obtained after DOI correction is 9.6 keV FWHM.}
\label{fig:spectra}
\end{figure}
Au as cathode material, and compared four different anode materials Au, Cr, In, and Ti
(see Fig.~\ref{fig:contactmaterial}). We measured the performance for the central 16 pixels, i.e. four
pixels of each material. We analyzed each pixel carefully and did a detailed 
analyses of the detection efficiency on the depth of interaction (DOI). 
One Ti pixel showed a drastic efficiency drop for a region in the middle 
of the detector and was excluded from the analyses. 
We obtain the best energy resolutions for the Ti  and In pixels with little 
variation between those two metals. For a Ti pixel we obtain 
59 keV, 122 keV and 662 keV energy resolutions of 1.4 keV (4.0 keV), 1.9 keV (4.2 keV), and 7.4 keV (9.6 keV), respectively (values in brackets are before subtraction of 
amplifier noise). Fig. \ref{fig:spectra} shows the 662 keV energy spectrum.
\begin{figure}[h]
\centering
\includegraphics[width=10.5cm]{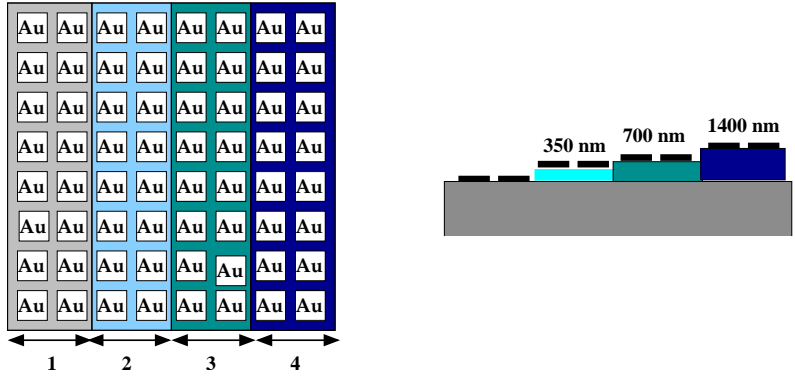}
\caption{We fabricated a detector to measure contact-contact IV-curves as function 
of the thickness of the Al$_2$O$_3$ isolation layer below the contacts. The left side 
shows a view of the anode side, the right side shows a cross-section (dimensions not to scale). 
Different colors mark layers with different Al$_2$O$_3$ thicknesses.}
\label{fig:MessungOxischicht}
\end{figure}
\section{Oxidation layer }
\label{sec:oxi}

The voltage that can be applied to a steering grid is limited by the current between the grid 
and the anode contacts. To overcome this problem, we explored the possibility to isolate contacts 
from the CZT substrate by a high-resistivity layer. The idea is to use this technique for 
the steering grid, see Fig. \ref{fig:Layer}. As isolation material we chose Al$_2$O$_3$ 
because of its high resistivity ($> 10^{14}\:\Omega\:$cm ) and excellent mechanical properties. 

In order to examine the effect of oxidation layers we deposited gold contacts on Al$_2$O$_3$ layers 
of different thicknesses (0~nm, 350~nm, 700~nm, 1400~nm) on a CZT substrate (see Fig. \ref{fig:MessungOxischicht}). We then measured the current between adjacent 
contacts as function of the applied bias. The thickest Al$_2$O$_3$ layer (1400~nm)
partially peeled off and contacts were deposited  on top of the 
Al$_2$O$_3$ and as well as directly on the CZT. For these contacts we 
did not measure IV curves.

\begin{figure}[b]
\centering
\includegraphics[width=8cm]{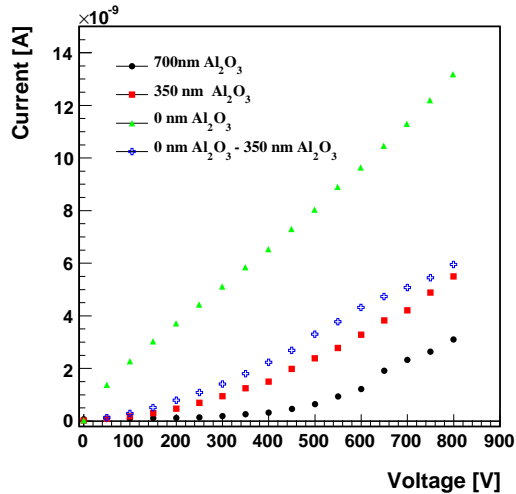}
\caption{The panel shows the IV-curves between adjacent gold contacts deposited
directly on the CZT or on Al$_2$O$_3$ layers of different thicknesses.
All contacts have the same pad area and the same spacing between them. 
If only one of the contacts is isolated by Al$_2$O$_3$, the IV-curves 
depend on the sign of the applied voltage. We only give the dependency for the contact with 0~nm Al$_2$O$_3$ at ground and negative voltages applied to the contact with 350~nm Al$_2$O$_3$. The graph shows that 
Al$_2$O$_3$ layers can substantially reduce contact-contact currents.}
\label{fig:CurrentRaumTemp}
\end{figure}

Fig. \ref{fig:CurrentRaumTemp} shows the IV curves between different contacts. 
It can be recognized that the Al$_2$O$_3$ layers do indeed reduce the currents between
adjacent contacts, even if one of the contacts is not isolated.
In the latter case, the IV curve is not symmetrical and the current depends 
on the sign of the applied voltage.   We only give the dependency for the contact with 0~nm Al$_2$O$_3$ at ground and negative voltages applied to the contact with 350~nm Al$_2$O$_3$. One can clearly see, that the thicker the isolating layer, the smaller 
the current at constant voltage. 

\begin{figure}
\centering
\includegraphics[width=16.5cm]{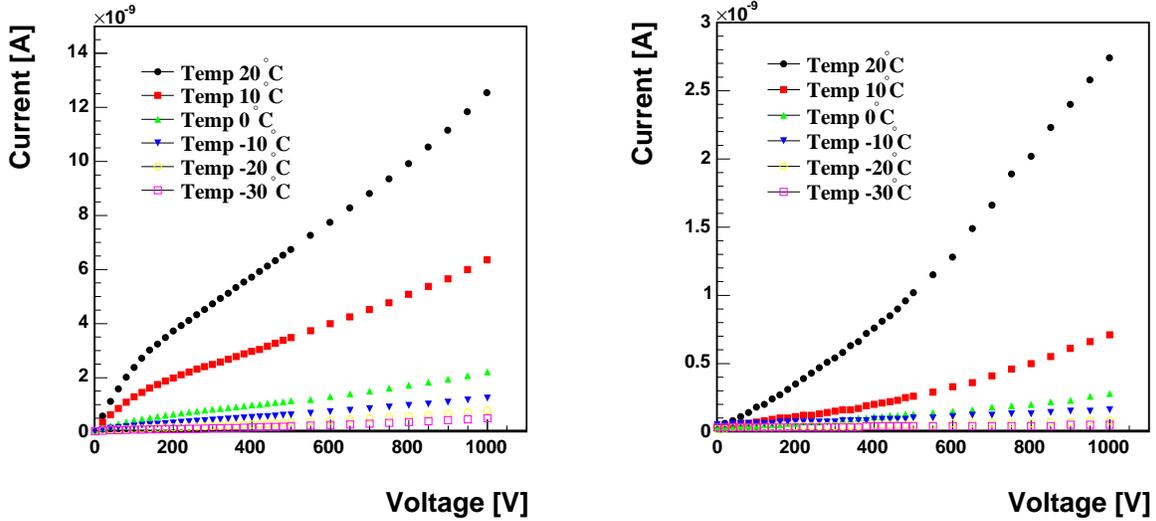}
\caption{The panels show contact-contact IV-curves for different temperatures.
In the left panel, the Au contacts were directly deposited on the CZT substrate.
In the right panel, both contacts were isolated from the CZT substrate by a 700~nm 
thick Al$_2$O$_3$ layer.}
\label{fig:OxiTemperatur}
\end{figure}

In satellite experiments like the proposed Energetic X-ray Imaging Survey Telescope 
({\it EXIST}) satellite \cite{Grindlay}, CZT detectors would 
be used at minus-Centigrade temperatures.
We thus measured the contact-contact IV curves for temperatures between 
-30$^{\circ}$ and +20$^{\circ}$ in 10$^{\circ}$ steps for two contacts
deposited directly on the CZT and for two contacts deposited on top of 
700~nm Al$_2$O$_3$. Fig. \ref{fig:OxiTemperatur} shows the results.
The currents decrease as the temperature decreases from 20$^{\circ}$ 
to -30$^{\circ}$: at 800~V bias, from 9.9~nA to 0.36~nA without isolation 
layer and from 2~nA to 0.04~nA with isolation layer.
Thus, lowering the temperature from 20$^{\circ}$ to -30$^{\circ}$ reduces 
the bias current by more than one order of magnitude even without any 
isolation layer, but by two orders of magnitude with the 700~nm  Al$_2$O$_3$ layer.
\section{Performance of a CZT detector with and without an isolated steering grid}
\begin{figure}[h] 
\centering
\includegraphics[width=6cm]{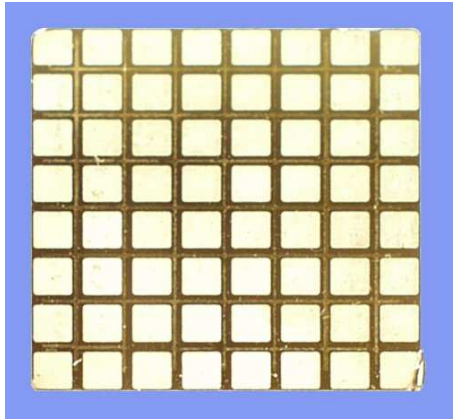}
\caption{Pixelated 2.0$\times$2.0$\times$0.5~cm$^3$ CZT detector  with steering grid. A high-resistivity Al$_2$O$_3$
film isolates the steering grid from the CZT substrate.}
\label{photo}
\end{figure}

We tested an Orbotech CZT detector (2 x 2 x 0.5 cm$^3$) contacted with an In cathode and 
In pixels before and after depositing an isolated steering grid. 
We used here a 150~nm thick Al$_2$O$_3$ layer. Although the results from the previous section 
show that the isolation layer should be thicker, we obtained encouraging results even 
with the thin isolation layer. The detector is shown in Fig.\ \ref{photo}.

\begin{figure}[b] 
\centering
\includegraphics[width=8cm]{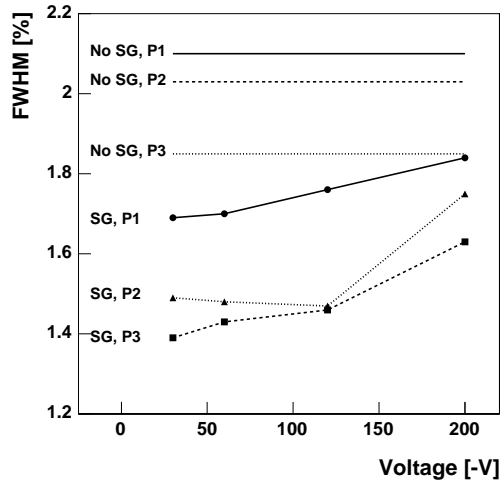}
\caption{Energy resolutions of the three central pixels of a detector contacted with an In cathode and
with In anode pixels. The horizontal lines show the energy resolutions without a steering grid (``No SG''), 
and lines labeled (``SG'') and plotted data points give the energy resolutions with an isolated 
steering grid at different bias voltages.}
\label{EnergyResolution}
\end{figure}

Figure \ref{EnergyResolution} shows the energy resolutions of the three central pixels 
for flood-illumination with 662 keV photons.  
Without a steering grid, the three pixels achieved energy resolutions of 2.10\%, 2.03\% 
and 1.85\%. With steering grid biased at -30~V, the performance of the same three 
pixels improved to 1.69\%, 1.49\% and 1.39\%, respectively.
The energy resolutions changed little for steering grid voltages between 
0~V and -200~V. Below -200~V, the performance deteriorates. 
We conclude (see section \ref{sec:oxi}) that the isolation was still too thin, 
as the grid-pixel current deteriorated the energy resolution. 
A detailed analysis of data taken with a collimated X-ray beam shows that 
the grid improves the detection efficiency for events between pixels.
The number of good events increases by a factor of 3 for events 
in the gap between pixels.

\begin{figure} 
\centering
\includegraphics[width=15cm]{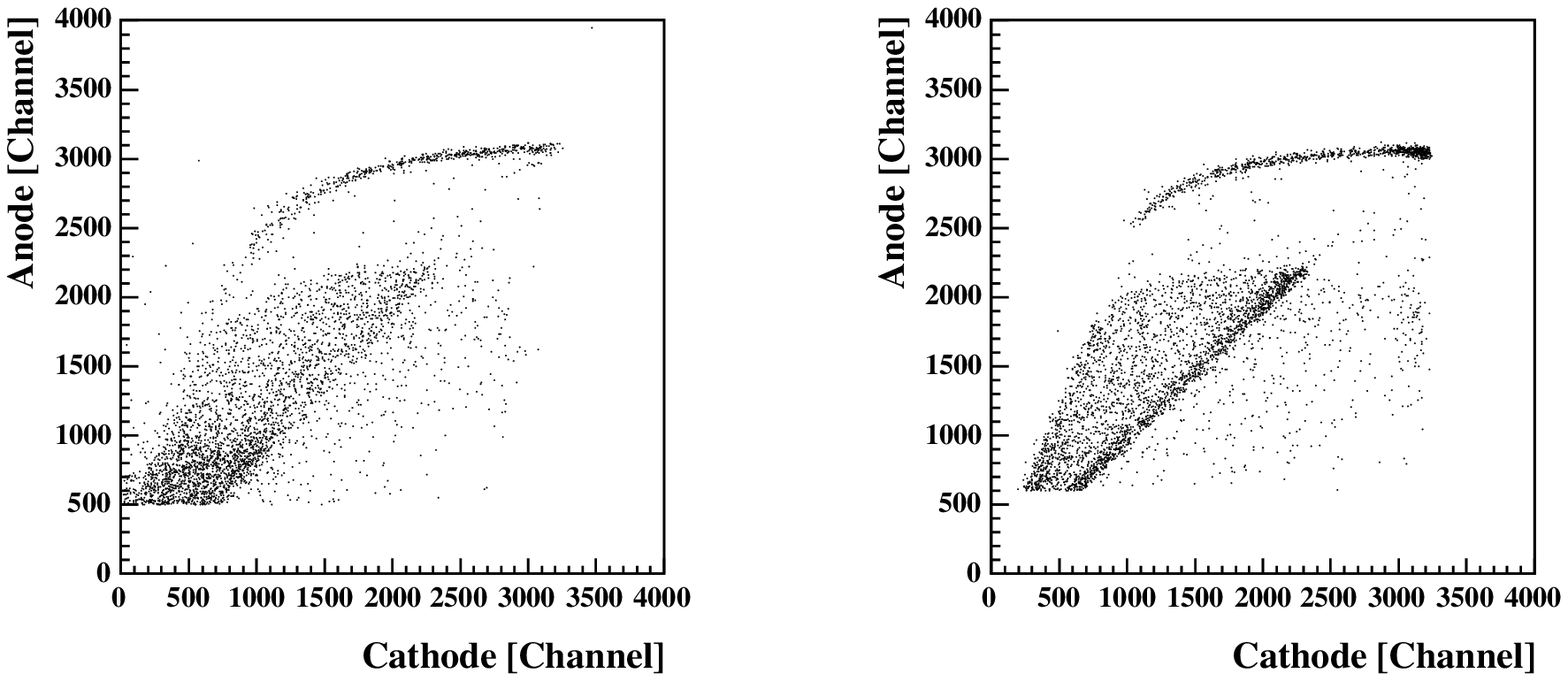}
\caption{The left panel shows the anode-cathode correlation for a detector 
with isolated steering grid  illuminated  with a collimated $^{137}$Cs 662 keV 
photon beam directed  towards the center of the pixel. 
The same correlation is shown  on the right side for simulations, taken into account an electron mobility of $\mu_e=720\:$cm$^2$~/V/s, electron lifetime 
of $\tau_e=8\cdot10^{-6}$~s, a hole mobility of $\mu_h=130\:$cm$^2$~/V/s and a hole 
lifetime of $\tau_h=1.5\cdot10^{-6}$~s.
 }
\label{fig:AnodevsCathode}
\end{figure}

At the end of this contribution, we want to show that detector simulations can be 
used to model the data from the CZT detector with astonishing success. 
We use the Geant 4 code for simulating the photon interactions, a 3-D Poisson solver 
developed by S. Komarov, and an in-house developed code for tracking electrons 
and holes through the detector. Figure \ref{fig:AnodevsCathode} shows the anode signal 
as a function of the cathode signal for experimentally measured and simulated 
data (662 keV flood illumination). 
The detector was radiated by a collimated $^{137}$Cs source. We get an excellent agreement 
when using an electron mobility of $\mu_e=720\:$cm$^2$~/V/s, electron lifetime 
of $\tau_e=8\cdot10^{-6}$~s, a hole mobility of $\mu_h=130\:$cm$^2$~/V/s and a hole 
lifetime of $\tau_h=1.5\cdot10^{-6}$~s. The electron mobility is consistent with 
an estimate based on the applied voltage and the measured electron drift times.
\section{Summary and Conclusions}
In this report, we present the results from several activities to optimize 
Orbotech CZT detectors. We tested detectors with standard In anode contacts and 
different cathode contacts. All the cathode contacts reduced the leakage 
current by more than one order of magnitude compared to the original In cathode. Au, Cr, In and Ti cathodes gave very similar 662 keV energy resolutions.
 On a different CZT detector, we replaced both, anode 
and the cathode contacts with identical material. In this study, Au gave the 
best results, followed in order of performance by Pt, In, and Cr.
We tested four different anode materials. Ti and In gave excellent almost identical results. 
For further improvements one can use steering grids. Our results show that one can use
an $\sim$700~nm thick Al$_2$O$_3$ isolation layer beneath the steering grid to 
reduce the current between grid and pixel. A first detector with isolated steering 
grid shows a significant improvement in energy resolution and detection efficiency.
We plan to test other isolation materials and to continue to optimize the 
use of Al$_2$O$_3$. One can use isolation layers also for other CZT detectors designs such as 
Frish-grid and cross-strip detectors.
%
\acknowledgments  
We are grateful to S.~Komarov for sharing his Poisson solver with us.
We thank Orbotech Inc. for their cooperation. We acknowledge L.~Sobotka, 
D.~Leopold, and J.~Buckley for helpful discussions. Thanks to electrical 
engineer P.~Dowkontt, and electrical technician G.~Simburger for their support.
This work is supported by NASA under contracts NNG04WC176 and 
NNG04GD70G, and the NSF/HRD grant no.\ 0420516 (CREST).
\bibliography{report}   

\begin{thebibliography}{1}
\bibitem{Orbotech}{Orbotech Inc., Rabin Park, 10 Plaut Street, Rehovot, Israel}
\bibitem{Vada}{Vadawale, S. V. et al., Proc. of SPIE 5540, p. 22-32, 2004}
\bibitem{Henric04}{ Krawczynski, H., Jung, I., Perkings, J., Burger, A., Groza, M.   Proc. of SPIE 592202-1, 2005}
\bibitem{Jung05}{ Jung, I., Groza, M., Perkins, J., Krawczynski, H., Burger, A.  Proc. of SPIE 592202-1, 2005}

\bibitem{Barret95}{Barret, H. H., Eskin, J. D. 1995, Phys. Rev. Lett., 75, 156}
\bibitem{Luke95}{Luke, P.N. 1995, In: Procs. Of the "9th International Workshop on Room Temperature Semiconductor X- and Gamma-Ray Detectors, Associated Electronics and Applications", Grenoble, France, 18-22 Sept., 1995}
\bibitem{Kalem02}{Kalemci, E., Matteson, J., Nucl. Instr. Methods in Phys Research Section A, 2002, 478: p. 527-537}
\bibitem{Zhang04}{Zhang, F., He, Z. Proceedings SPIE, p. 135-143, 5540, 2004 }
\bibitem{Kraw04}{Krawczynski, H.,  Jung, I., Perkins, J., Burger, A., Groza, A. Proceedings SPIE, p. 49, 5540, 2004}
\bibitem{bolotnikov}{Bolotnikov, A., Nucl. Instr. Methods in Phys Research Section A, 1999, 432: p. 326-331}
\bibitem{shipley}{Shipley Company, LLC, 455 Forest Street, Marlborough, MA 01752, USA}
\bibitem{G4}
S.~Agostinelli et al., \emph{Geant4--a simulation toolkit}, NIM A506, 250-303,2003
\bibitem{Grindlay}{Grindlay, J. E. \& the EXIST Team 2005, New Astron. Rev., 49, 435, 2005}
\bibitem{Jung:06}{Jung, I. Astroparticle Physics, in press}
\end{thebibliography}
\bibliographystyle{spiebib}   

\end{document}